\documentclass[amsmath,amssymb,superscriptaddress,twocolumn, showkeys, prl]{revtex4-1}

\usepackage{graphicx}
\usepackage{dcolumn}
\usepackage{bm}
\graphicspath{{snrfigs/}}

\begin{document}

\title{Dynamics of cooperativity in chemical sensing among cell-surface receptors}

\author{Monica Skoge} \affiliation{Department of Biology, University
  of California San Diego, San Diego, CA 92037}

\author{Yigal Meir}
\affiliation{Department of Physics, Ben-Gurion University,
Beer Sheva, Israel 84105}

\author{Ned S. Wingreen}
\affiliation{Department of Molecular Biology, Princeton University,
Princeton, NJ 08544}

\begin{abstract}
Cooperative interactions among sensory receptors provide a general
mechanism to increase the sensitivity of signal transduction.  In
particular, bacterial chemotaxis receptors interact cooperatively to
produce an ultrasensitive response to chemoeffector concentrations.
However, cooperativity between receptors in large macromolecular
complexes is necessarily based on local interactions and consequently
is fundamentally connected to slowing of receptor conformational
dynamics, which increases intrinsic noise.  Therefore, it is not clear
whether or under what conditions cooperativity actually increases the
precision of the concentration measurement.  We explictly calculate
the signal-to-noise ratio (SNR) for sensing a concentration change
using a simple, Ising-type model of receptor-receptor interactions,
generalized via scaling arguments, and find that the optimal SNR is
always achieved by independent receptors.
\end{abstract}

\maketitle

In biological networks, cooperative interactions among components can
sharpen input-output relations, increasing gain and enabling
switch-like responses.  The best-known example is the cooperative
binding/release of oxygen by hemoglobin, which enables efficient
transport of oxygen between the lungs and tissue. In sensory systems,
a well-studied example of cooperativity is receptor-receptor coupling
in {\it Escherichia coli} chemotaxis~\cite{SourjikBerg2002a}.  For
sensing systems, there is an obvious advantage of high gain to amplify
weak signals~\cite{Bray1998}, particularly when combined with an
adaptation system to broaden the dynamic range~\cite{Hansen2008}.
This advantage of high gain from receptor-receptor interactions raises
the question: why has receptor coupling not evolved in other chemical
sensing systems, e.g. quorum sensing and eukaryotic chemotaxis?

If the function of a sensory network is to reliably detect weak
signals, then processing of the signal is only half the story.  The
other half is the suppression of noise and for weak signals the
signal-to-noise ratio (SNR) generally governs information transmission
and sensory performance~\cite{suppinfo}.  Cellular signal transduction
must contend with both noisy inputs (extrinsic noise), as well as the
noise generated internally by the signal transduction system itself
(intrinsic noise).  While the presence of signaling noise is well
appreciated, the connection between cooperativity and noise has
received less attention~\cite{Berg2000, *ShibataFujimoto2004,
  BialekSetayeshgar2008, *Aquino2011, *Hu2010}.  As we will show,
receptor cooperativity and intrinsic signaling noise are inextricably
linked via the statistical mechanics of receptors.  The same
cooperative interactions that give rise to enhanced sensitivity
necessarily both amplify fluctuations and slow the rate of
receptor-conformational switching, limiting both the response time and
the ability of the system to reduce intrinsic noise by time-averaging.
Due to these tradeoffs, it is unclear when or whether receptor
cooperativity actually increases sensory performance.

The prevailing view of bacterial chemoreceptor operation is that
signal amplification from receptor cooperativity enables the
chemotaxis network to reliably detect shallow gradients with exquisite
sensitivity.  In this Letter, we challenge this viewpoint by
evaluating the strategy of using receptor cooperativity to enhance
weak signal detection in light of the tradeoffs between gain and
intrinsic noise.  Specifically, we calculate the SNR for a simple
physical model of receptor-receptor interactions, a dynamical
Ising-type model in which receptors have two conformational states,
active and inactive, and neighboring receptors prefer to be in the
same conformational state (a.k.a. conformational
spread~\cite{Bray1998}).  The input to the network is the (changing)
external ligand concentration and the output is the time-averaged
total receptor activity.  This two-state model is a good description
of bacterial chemoreceptors, in which the active and inactive
conformational states are named for their relative abilities to
activate a downstream kinase~\cite{WadhamsArmitage2004} and the output
of the network is the CheY-P concentration, which is effectively the
time-average of receptor activity over the phosphorylated lifetime of
individual CheY-P molecules.

Importantly, our model is the first to both (i) incorporate the {\em
  dynamics} of receptor switching, which is key to understanding the
reduction in noise by time-averaging and (ii) base cooperativity on
{\em local} interactions.  Unlike the dynamic MWC models used in
previous work~\cite{BialekSetayeshgar2008, *Aquino2011}, we invoke no
unphysical action-at-a-distance that allows the entire receptor
cluster to flip conformational states simultaneously.  In fact, a
recent study has provided direct evidence for local interactions
underlying the ultrasensitive behavior of the bacterial
motor~\cite{Bai2010}.

We consider a cell with a total of $N$ two-state receptors, which are
divided into $m$ independent 1D chains each of length $n$, with
nearest-neighbor Ising couplings inside each chain, given by the
Hamiltonian
\begin{equation}
H = -J \sum_{\langle i,j \rangle} \sigma_i \sigma_j + \frac{1}{2}\sum_{i=1}^n
f\sigma_i,
\end{equation}
where $\sigma_i=\pm 1$ represents the active/inactive receptor states,
$\langle i,j \rangle$ denotes nearest neighbors, and 
\begin{equation}
f = \Delta \epsilon +
\log{\left(\frac{1+[\text{L}]/K^{\text{off}}}{1+[\text{L}]/K^{\text{on}}}
\right)}
\label{f}
\end{equation}
is the free-energy difference between active and inactive states,
which is a function of the ligand concentration $[L]$, ligand
dissociation constants $K^{\text{off}}$ and $K^{\text{on}}$, and the
``offset energy'' $\Delta \epsilon \equiv E_{\text{on}} -
E_{\text{off}}$, which is the energy difference between active and
inactive states in the absence of
ligand~\cite{Keymer2006}~\footnote{All energies are in units of the
  thermal energy $k_B T$.}~\footnote{Note that for finite $n$ in the
  limit $J\rightarrow \infty$, our Ising model of receptor coupling
  becomes the ``all-or-none'' Monod-Wyman-Changeux (MWC)
  model~\cite{Monod1965}.}.

When ligand binding/unbinding is much faster than receptor
conformational switching, the stochastic dynamics of switching is
governed by a kinetic Ising model, where the occupation probablity
$p(\sigma_1, \dots \sigma_n; t)$ obeys the master
equation
\begin{eqnarray}
\frac{d}{dt} p(\sigma_1, \dots \sigma_n; t) & = & -\left(\sum_i
k_i(\sigma_i)\right)p(\sigma_1, \dots \sigma_n; t)\label{master} \\ &
& + \sum_i k_i(-\sigma_i)p(\sigma_1, \dots, -\sigma_i, \dots \sigma_n;
t).\nonumber
\end{eqnarray}
(For the case of slow ligand dynamics, see~\cite{suppinfo}.)  We
assume the switching rates $k$ obey detailed balance and are local,
{\it i.e.} depend only on the conformational states of the receptor
and its nearest neighbors.  Although the form of this dependence has
not been measured experimentally, a simple, physically reasonable
choice is Glauber dynamics~\cite{Glauber1963}, which models
interactions of the receptor system with a heat bath.  With this
choice,
\begin{equation}
k_i(\sigma_i) = \frac{1}{2}\alpha \left(1 - \frac{1}{2}\gamma
\sigma_i(\sigma_{i-1} + \sigma_{i+1})\right)(1 - \beta \sigma_i), \label{glauberrates}
\end{equation}
where $\alpha$ sets the intrinsic switching rate (typically
$~10^{3}-10^{4}/$s), $\gamma = \tanh(2J)$, and $\beta =
-\tanh(f/2)$~\footnote{Note that these rates obey detailed balance and
  the minimal and maximal flipping rates are $0$ and $\alpha$,
  respectively.  As such, a large free-energy difference can make the
  rate of switching from a low-energy state to a high-energy state
  arbitrarily small, but the reverse transition can only approach the
  maximal rate $\alpha$}.

We let the model system respond to a small step in ligand
concentration (and associated step change in free energy $\Delta
f\sim\Delta \log{[\text{L}]}$) over a given period of measurement,
which we call $\tau_{\text{avg}}$.  We define the average activity of
a receptor to be the probability the receptor is in the active state,
with the total activity $A$ being the sum of the individual receptor
activities.  The output of the system is the time-averaged change in
total activity~\footnote{This output is sensible from
  a biological point of view, {\em e.g.} an internal protein pool of
  activated protein reflects the output of the receptors over an
  averaging time set by the turnover time of the activated proteins.},
\begin{eqnarray}
\Delta A(\tau_{\text{avg}}) & = & \frac{1}{\tau_{\text{avg}}} \int_0^{\tau_{\text{avg}}} \int_0^{t}\chi(t-t') \Delta f(t')dt' \\
& \equiv & \frac{1}{4}R(\tau_{\text{avg}}) \Delta f, \label{responseeq}
\end{eqnarray}
where the dynamical susceptibility $\chi$ relates changes in the
average cluster activity to time-dependent changes in the free-energy
difference and Eq.~\ref{responseeq} defines the response function
$R(\tau_{\text{avg}})$.  The system is assumed to be pre-adapted to
the ambient ligand concentration, such that the free-energy difference
between active and inactive states is zero prior to stimulation ({\it
  i.e.} the system is adapted to the most sensitive region of the
input-output relation); this assumption appears to be correct for
bacterial chemoreceptors~\cite{Keymer2006}.

\begin{figure}
\begin{center}
\includegraphics*[scale=0.24,angle=0]{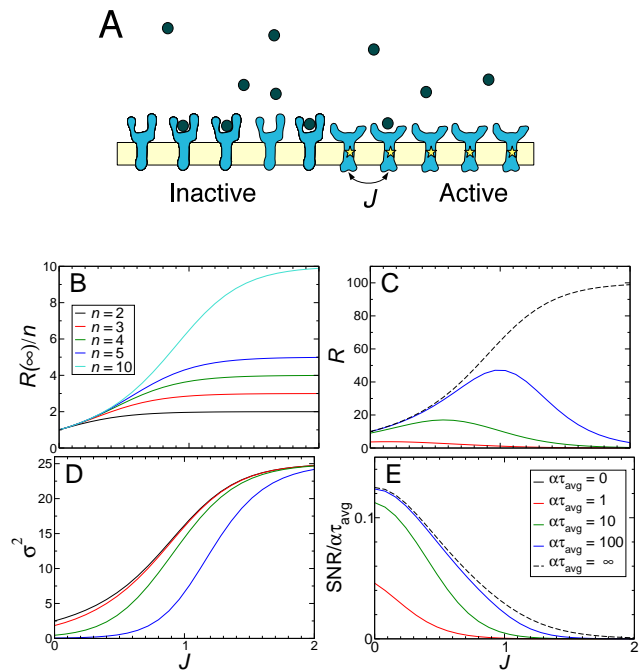}
\\
\includegraphics*[scale=0.16,angle=0]{figure1b.eps}
\includegraphics*[scale=0.16,angle=0]{figure1c.eps}
\\
\includegraphics*[scale=0.16,angle=0]{figure1d.eps}
\includegraphics*[scale=0.16,angle=0]{figure1e.eps}
\caption{(A) Schematic diagram of cooperatively interacting receptors.
  Receptors interconvert between active and inactive states and
  preferentially bind ligand in the inactive state.  The interaction
  energy $J$ between neighbors favors receptors in the same activity
  state.  (B) The normalized static response $R(\infty)/n$ for closed
  chains (rings) of various lengths $n$ (for open chains,
  see~\cite{suppinfo}).  (C) Response $R$, (D) noise $\sigma^2$, and
  (E) SNR as functions of coupling strength $J$ for a closed chain of
  $n=10$ receptors for various averaging times $\tau_{\text{avg}}$.
  In (E), the SNR is normalized by the averaging time.  }
\label{figure1}
\end{center}
\end{figure}

The sensitivity to signal is quantified by the static response
$R(\infty) = 4\int_0^{\infty} \chi(t)dt$, which, when normalized by
chain length, is the factor by which the DC ({\it i.e.} infinite time)
response of a coupled receptor is amplified relative to that of an
uncoupled receptor.  The normalized static response $R(\infty)/n$
increases as a function of coupling strength from $1$ to $n$, as shown
in Fig.~\ref{figure1}B.  At zero coupling, $J=0$, each receptor
behaves independently and there is no cooperative amplification of the
signal.  As the coupling is increased, domains of adjacent receptors
begin to effectively switch conformations together and the
amplification $R(\infty)/n$ is determined by the size of these
domains.

In practice, many cells have a limited measurement period and this can
decrease the response $R(\tau_{\text{avg}})$, as shown in
Fig.~\ref{figure1}C.  For a 1D ring, this dependence takes the simple
form
\begin{equation}
R_{\textrm{ring}}(\tau_{\text{avg}}) =  R_{\textrm{ring}}(\infty)\left(1-\frac{\tau^{\textrm{ring}}_c}{\tau_{\text{avg}}}\left(1-e^{-\tau_{\text{avg}}/\tau^{\textrm{ring}}_c}\right)\right)
\end{equation}
where 
\begin{equation}
\tau^{\textrm{ring}}_c = 1/(\alpha(1-\tanh(2J)))  \label{tauc} 
\end{equation}
is the response time, which increases exponentially with coupling
strength due to the well-known phenomenon of critical slowing
down~\cite{hohenberg}.  Notice that for any finite averaging time, the
response eventually falls to zero with increasing coupling strength
because the system slows down and cannot respond to the input in the
time available.  Intuitively, for large coupling $J$, the receptors
become frozen in an all-active or all-inactive state, with switching
between these states too slow to mediate a timely response to a
changing input.  

The intrinsic noise, {\it i.e.} the variance in time-averaged
activity, increases monotonically with coupling strength, as shown in
Fig.~\ref{figure1}D.  For the 1D ring,
\begin{equation}
\sigma_{\textrm{ring}}^2(\tau_{\text{avg}}) =  \sigma^2_{\textrm{ring}}(0)\mathcal T
(\tau_{\text{avg}}/\tau^{\textrm{ring}}_c), \label{noisering}
\end{equation}
where $\tau_c$ is given in Eq.~\ref{tauc} and 
\begin{equation}
{\mathcal T}(x) = 2(x + e^{-x}-1)/x^2.  
\end{equation}  
For a given $J$, the maximal value of the noise occurs for zero
averaging time (the ``snapshot'' limit), and is proportional to the
static response as required by the fluctuation-dissipation theorem.
Averaging for longer times substantially reduces the noise (via the
factor $\mathcal T (\tau_{\text{avg}}/\tau^{\textrm{ring}}_c)$), but
time-averaging becomes less effective with increasing $J$ due again to
critical slowing down, as the slower dynamics increases the
correlation time $\tau_c$, the time required for fluctuations in
activity to decay.

The relative uncertainty in sensing the concentration of ligand is
determined by the signal-to-noise ratio (SNR), as shown in
Fig.~\ref{figure1}E, normalized per receptor,
\begin{equation}
\text{SNR} \equiv \frac{\text{SNR}_{\text{total}}}{N(\Delta f)^2} =
\frac{(m\Delta A(\tau_{\text{avg}}))^2}{m\sigma^2(\tau_{\text{avg}})N(\Delta f)^2}
= \frac{R^2(\tau_{\text{avg}})}{16n
  \sigma^2(\tau_{\text{avg}})}.\label{SNR}
\end{equation}
Surprisingly, the optimal value of the coupling strength $J$ to
maximize the SNR is zero for all averaging times.  That is, on a per
receptor basis, independent receptors always have a lower total SNR
than cooperative teams of receptors.  For short averaging times,
independent receptors have both a larger response and lower noise than
teams.  For longer averaging times, cooperative teams have a larger
response, but independent receptors still achieve a higher SNR because
their rapid switching leads to more independent samples of receptor
activity and thus to a much lower time-averaged noise.  

While the results in Fig.~\ref{figure1}C-E are shown only for $n=10$,
scaling analysis indicates that increasing the chain length $n$ will
never make cooperative teams favorable with respect to uncoupled
receptors.  In the following, we derive how the SNR scales with the
only two length scales in our model: the chain length $n$ and the
correlation length $\xi$, which is the length scale over which the
conformational states of neighboring receptors are correlated in an
infinite system.  

To best realize the cooperativity of a 1D chain of length $n$, the
coupling strength has to be set to $J^{*}_{n}$, the smallest $J$ that
gives approximately maximal response ($R(\infty) \approx n^2$), and
the averaging time has to be just long enough for
$R(\tau_{\text{avg}})$ to approach this maximum, {\it i.e.}
$\tau_{\text{avg}} \approx \tau_c(J^{*}_{n})$.  Longer times will
reduce the noise, but by exactly the same factor for both the
cooperative team and independent receptors, $\propto
1/\tau_{\text{avg}}$.  

For the cooperative team, $\tau_{\text{avg}}=\tau_c(J_n^*)$ is long enough that
$R(\tau_{\text{avg}})\approx R(\infty)$, while $\sigma^2(\tau_{\text{avg}}) \sim
\sigma^2(0)$ ({\it i.e.}  still roughly in the snapshot limit for
noise).  By the fluctuation-dissipation theorem,
$\sigma^2(0)=R(\infty)/4$, so 
\begin{equation}
\text{SNR}(J^{*}_n) \approx \frac{R^2(\infty)}{16n \sigma^2(0)} = \frac{R(\infty)}{4n} \approx n/4 \approx \xi(J^{*}_{n})/4.\label{SNRteam}
\end{equation}
Thus the SNR for the cooperative team is approximately equal to the
chain length.  The last approximate equality holds because the $J^*_n$
required for nearly maximal response yields an (infinite-chain)
correlation length $\xi(J^{*}_{n})$ comparable to the actual chain
length $n$.

By comparison, for $n$ independent receptors, the same averaging time
$\tau_{\text{avg}} = \tau_c(J^*_n)$ is long enough to reach the static
response $R(\tau_{\text{avg}})\approx R(\infty) = n$, and for substantial noise
reduction by time-averaging, $\sigma^2(\tau_{\text{avg}}) \approx
2\sigma^2(0)/(\alpha \tau_{\text{avg}})$, so that 
\begin{eqnarray}
\text{SNR}(J=0) & \approx & \frac{R^2(\infty)}{32n \sigma^2(0)/(\alpha \tau_{\text{avg}})} =
\frac{n^2}{8n^2/(\alpha \tau_{\text{avg}})}\nonumber \\ & = & \frac{1}{8}\alpha
\tau_{\text{avg}} \approx \frac{1}{8} \alpha \tau_c(J^{*}_{n}).\label{SNRindep}
\end{eqnarray}

\begin{figure}
  \begin{center}
    \includegraphics*[scale=0.25,angle=0]{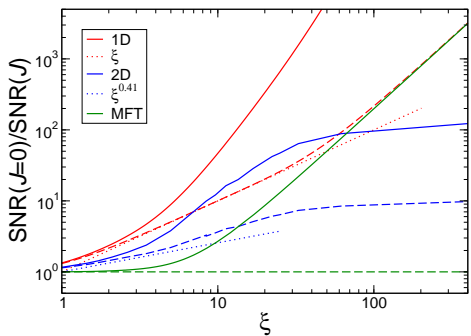}
    \caption{Ratio of the SNR for independent receptors to the SNR for
      cooperative receptors as a function of the correlation length
      $\xi$ for a 1D ring and a 2D square lattice of $n=100$ receptors
      with periodic boundary conditions, as well as mean field
      theory (MFT).  Averaging times are $\alpha
      \tau_{\text{avg}}=100$ (solid curve) and $\alpha \tau_{\text{avg}}=\infty$
      (dashed curve), with power law plots for comparison (dotted
      lines).  }
    \label{SNRratiofig}
  \end{center}
\end{figure}

Therefore, for cooperative teams to be favorable with respect to uncoupled
receptors requires $\xi(J^{*}_{n})>\frac{1}{2}\alpha
\tau_c(J^{*}_{n})$, according to Eqs.~\ref{SNRteam}
and~\ref{SNRindep}.  However, this will never happen for 1D chains
because the correlation time grows faster than the correlation length
with increasing $J$ in 1D, specifically $\tau_c\sim\xi^2$.  In
Fig.~\ref{SNRratiofig} we plot the ratio of SNRs for independent
receptors versus cooperative receptors for closed chains.  
The ratio never favors cooperative receptors and, in the
long-averaging-time limit, follows the expected scaling $\sim \xi$ for
$\xi\ll n$ and $\sim \xi^2$ for $\xi\gg n$.  A shorter averaging time
only makes things worse for cooperative receptors, as in this case the
expected scaling is still $\sim \xi$ for $\xi\ll n$, but $\sim \xi^4$
for $\xi\gg n$~\cite{suppinfo}.

Our results naturally generalize to higher-dimensional coupling.  For
2D, the most natural topology for interacting membrane receptors, the
ratio of uncoupled to coupled SNRs for long averaging times
$\tau_{\text{avg}}\gg\tau_c$ scales as
$\text{SNR}(J=0)/\text{SNR}(J)\rightarrow n\alpha \tau_c/R(\infty)
\sim \xi^{z-\gamma/\nu} \approx
\xi^{0.41}$~\cite{suppinfo}~\footnote{Exponents have their
  conventional definitions, $\xi\sim|T-T_c|^{-\nu}$,
  $\chi\sim|T-T_c|^{-\gamma}$, and $\tau_c\sim\xi^z$}, as seen in
Fig.~\ref{SNRratiofig} in the expected regime $\xi\ll \sqrt{n}$.  As
in 1D, independent receptors do better than coupled receptors, but
this advantage grows more slowly with correlation length in 2D.

High dimensional ($D\geq4$), or global coupling of receptors, which
could be mediated by the cell membrane or by rapidly diffusing
effectors, is described by the mean field limit of the Ising model.
In this limit, the dynamics of cluster activity reduces to that of an
overdamped harmonic oscillator, and the normalized static response and
correlation time are both equal, $R^{\text{MFT}}(\infty)/n =
\alpha\tau_c^{\text{MFT}} = 1/(1-\zeta J)$, where $\zeta$ is the
number of nearest neighbors.  Consequently, for long averaging times,
the ratio of SNRs for independent versus coupled receptors is
constant.  Thus, at best, increasing the dimension of receptor
coupling yields parity between cooperative and independent receptors.

More generally, the universal behavior of Ising models near the phase
transition and a rigorous bound on Ising critical exponents $\nu z
\geq \gamma$~\cite{AbeHatano1969} suggest that critical slowing down
is unavoidable and independent receptors will {\it always} optimize
the SNR.  We have also studied the addition of other potential noise
sources, including slow ligand dynamics and static variation of
receptor offset energies, and find that independent receptors still
yield the best SNR~\cite{suppinfo}.

In summary, we developed a physical description of cooperativity based
on the principle that allosteric interactions between receptor
proteins are inherently local.  From our simple Ising-type model,
which encompasses a broad class of models, we elucidated the
relationship between cooperativity and intrinsic noise.  We found that
the slowing down of receptor switching due to cooperative interactions
strongly impairs the SNR and that consequently the SNR is always
highest for {\emph zero} receptor cooperativity, even though the
absolute sensitivity is optimized for nonzero cooperativity.  Since
our SNR is normalized by receptor number and stimulus strength, our
results show that (i) for a given small stimulus, independent
receptors achieve SNR $>1$ for the fewest receptors, and (ii) for a
given number of receptors, independent receptors achieve SNR $>1$ for
the smallest stimulus strength.

Our surprising result offers a fresh perspective on
bacterial chemotaxis, by indicating that the network is not simply
optimizing SNR, since the observed receptor cooperativity in this
system reduces the SNR by a factor of $\sim50$~\footnote{The observed
  cooperativity of {\it E. coli} chemotaxis receptors suggests that
  the receptors form strongly-coupled MWC clusters of size
  $n\approx10$~\cite{Keymer2006}.  We estimate $J$ for MWC clusters as
  the coupling required to give $95\%$ maximal response, which gives
  $J=2.4$ and $1.0$ for 1D ($n=10$) and 2D ($n=9$) open clusters,
  respectively, corresponding to a reduction in SNR (relative to
  independent receptors) by factors of $56$ and $50$, respectively.}.
More generally, our result reveals that the benefits of cooperativity
for sensing are far from obvious, potentially explaining the absence
of receptor cooperativity in many sensory networks.

\begin{acknowledgements}
We thank Edward Cox, Nathaniel Ferraro, Herbert Levine, and Anirvan
Sengupta for helpful discussions. This research was partially
supported by NIH grant 1 R01 GM078591 (M.S.), R01 GM082938 (Y.M. and
N.S.W.)  and by NSF Grant PHY-0957573 (N.S.W.).
\end{acknowledgements}

\bibliography{bibliography}

\end{document}